\documentclass{aastex}
\usepackage{emulateapj5}

\def\wig#1{\mathrel{\hbox{\hbox to 0pt{%
          \lower.5ex\hbox{$\sim$}\hss}\raise.4ex\hbox{$#1$}}}}

\shorttitle{Atmosphere, Interior, and Evolution of HD 149026b}
\shortauthors{Fortney et al.}

\newcommand{\mj}{$M_{\mathrm{J}}$}
\newcommand{\rj}{$R_{\mathrm{J}}$}
\newcommand{\me}{$M_{\oplus}$}
\newcommand{\T}{TrES-1}
\newcommand{\hd}{HD 209458b} 
\newcommand{\hh}{HD 149026b}
\newcommand{\he}{HD 189733b}
\newcommand{\cp}{\citep}
\newcommand{\ct}{\citet}

\slugcomment{\bf}
\slugcomment{Accepted to ApJ, December 31, 2005}

\begin{document}

\title{Atmosphere, Interior, and Evolution\\ of the Metal-Rich Transiting Planet HD 149026b}

\author{J. J. Fortney\altaffilmark{1}, D. Saumon\altaffilmark{2}, M. S. Marley\altaffilmark{1}, K. Lodders\altaffilmark{3}, R. S. Freedman\altaffilmark{1}$^,$\altaffilmark{4}}

\altaffiltext{1}{Space Science and Astrobiology Division, NASA Ames Research Center, MS 245-3, Moffett Field, CA 94035; jfortney@arc.nasa.gov}
\altaffiltext{2}{Los Alamos National Laboratory, MS P365, Los Alamos, NM 87545}
\altaffiltext{3}{Planetary Chemistry Laboratory, Department of Earth and Planetary Sciences, Washington University, St. Louis, MO 63130}
\altaffiltext{4}{SETI Institute, 515 N. Whisman Road, Mountain View, CA 94043}

\begin{abstract}

We investigate the atmosphere and interior of the new transiting planet \hh, which appears to be very rich in heavy elements.  We first compute model atmospheres at metallicities ranging from solar to ten times solar, and show how for cases with high metallicity or inefficient redistribution of energy from the day side, the planet may develop a hot stratosphere due to absorption of stellar flux by TiO and VO.  The spectra predicted by these models are very different than cooler atmosphere models without stratospheres.  The spectral effects are potentially detectable with the \emph{Spitzer Space Telescope}.  In addition the models with hot stratospheres lead to a large limb brightening, rather than darkening.  We compare the atmosphere of \hh~to other well-known transiting planets, including the recently discovered HD 189733b, which we show have planet-to-star flux ratios twice that of HD 209458 and \T.  The methane abundance in the atmosphere of \he~is a sensitive indicator of atmospheric temperature and metallicity and can be constrained with \emph{Spitzer} IRAC observations.  We then turn to interior studies of \hh~and use a grid of self-consistent model atmospheres and high-pressure equations of state for all components to compute thermal evolution models of the planet.  We estimate that the mass of heavy elements within the planet is in the range of 60 to 93 \me.  Finally, we discuss trends in the radii of transiting planets with metallicity in light of this new member of the class.

\end{abstract}

\keywords{planetary systems, radiative transfer, binaries:eclipsing, stars: individual (HD 149026, HD 189733)}


\section{Introduction}
Searches for extrasolar giant planets (EGPs) over the past ten years have yielded a bizarre menagerie of planets, many of which are in orbits with periods of only a few days.  These ``hot Jupiters,'' or ``Pegasi planets,'' which orbit within $\sim$0.1 AU of their parent stars, form a class of astrophysical objects that is still yielding surprises.  The latest, and perhaps most unanticipated, is \hh, a planet in a 2.87 day orbit around a metal-rich G0 IV parent star.  This planet was discovered by the radial velocity method and subsequently shown to transit its parent star \citep{Sato05}.  The most striking feature of planet \hh~is its small radius of only $0.725 \pm 0.05$ \rj, given its mass of $0.36 \pm 0.03$ \mj~(114 \me).  For comparison, Saturn has a mass of 0.30 \mj~(95 \me), is about 2.5 Gyr older than \hh, and receives 140,000 times less stellar flux, and yet has a mean radius of 0.81 \rj.  The transiting planet most similar in mass to \hh, OGLE-Tr-111b, has a mass of 0.53 \mj~and a radius of $\sim$1 \rj~\citep{Pont04}.

Evolution models of solar composition Pegasi planets have shown that lower-mass giant planets should be preferentially influenced by irradiation, leading to larger radii \citep{Guillot96}, but \hh~is decidedly small.  Recent interior structure models of Saturn show that it globally possesses $20\pm8$ \me~of heavy elements (including the core) for a corresponding 6--14 fold global enhancement in ``metals'' relative to the Sun \citep{Saumon04}.  The small radius of \hh~implies that it has even more heavy elements in its interior.  Indeed \citet{Sato05} have estimated a core mass near 70 \me, or 60\% of the planet's mass!  The high metallicity of the parent star [Fe/H]=0.36$\pm$0.05 \citep{Sato05} suggests that this system is very favorable to planet formation as the current fraction of stars with detected planets increases strongly with stellar metallicity \citep{Gonzalez97,Santos04,Fischer05}.

While the planet's orbital radius of 0.042 AU is similar to that of the well-known transiting planet \hd, the luminosity of the star HD 149026 is larger (2.7 $L_\odot$), and the degree of irradiation is intermediate between those of well known transiting planets OGLE-Tr-56b ($T_{\rm eff}\sim1950\,$K) and \hd~($T_{\rm eff}\sim1450\,$K).  We will show that the atmosphere of \hh~could be substantially different from that of \hd~and \T, which were recently detected in the infrared with the \emph{Spitzer Space Telescope} \citep{Deming05b,Charb05}.  As the HD 149026 system is relatively bright ($V=8.15$) it is an enticing target for follow up \emph{Spitzer} observations.

In this paper we investigate atmosphere models for the planet and show how under certain circumstances the atmosphere may feature a hot stratosphere driven by absorption of stellar flux by TiO and VO at millibar pressures.  We provide model planet-to-star flux density ratios in \emph{Spitzer} IRAC and MIPS bandpasses that will allow for a determination of the character of the planet's atmosphere.  We briefly discuss the atmospheres of all potential \emph{Spitzer} secondary eclipse targets, including the newly discovered \he~\citep{Bouchy05}.  We also investigate the evolutionary history of \hh~with the help of a new grid of model atmospheres computed specifically for this planet.  This atmosphere grid, applied to a planetary evolution code, with high-pressure equations of state of H/He mixtures, rock, and ice, allows us to place constraints on the abundances of heavy elements in \hh, and to begin to understand its current structure.

\section{Model Atmospheres}
\subsection{Methods}
Atmospheric pressure-temperature (\emph{P--T}) profiles and spectra for the planet were obtained with a plane-parallel model atmosphere code that has been used for a variety of planetary and substellar objects.  The code was first used to generate profiles and spectra for Titan's atmosphere by \citet{Mckay89}.  It was significantly revised to model the atmospheres of brown dwarfs \citep{Marley96, Burrows97, Marley02}, Uranus \citep{MM99}, cool EGPs \citep{Marley98}, and Pegasi planets \citep{Fortney05}.  It explicitly includes both incident radiation from the parent star and thermal radiation from the planet's atmosphere and interior.  The basic radiative transfer solving scheme was developed by \citet{Toon89}.  We use the elemental abundance data of \citet{Lodders03} and compute chemical equilibrium compositions with the CONDOR code, following \citet{Fegley94}, \citet{Lodders02}, and \citet{Lodders02b}.  Recently, \ct{Visscher06} have shown that while photochemical reactions likely will be important in Pegasi planet atmospheres at low pressure, chemical equilibrium abundances hold at pressures greater than $\sim$10 mbar.  For the most part, the infrared spectra of Pegasi planets are sensitive to opacity at $10 \lesssim P \lesssim 100$ mbar \citep{Fortney05}, so equilibrium chemistry calculations are probably sufficient.  This may need to be revisited as more observational data become available.

In addition we maintain a large and constantly updated opacity database.  Our dataset includes the molecular bands of H$_2$O, CH$_4$, CH$_3$D, CO, NH$_3$, H$_2$S, PH$_3$, TiO, VO, CrH, FeH, CO$_2$, HCN, C$_2$H$_2$, C$_2$H$_4$, and C$_2$H$_6$ complemented with the atomic lines of the alkali metals (Li, Na, K, Rb and Cs) from \citet{BMS} and continuum opacity sources from H$_2$ CIA, H$_2$, H and He Rayleigh scattering, and H$^-$ bound-free and free-free, H$_2^-$ free-free, He$^-$ free-free, and H$_2^+$ bound-free and free-free (Freedman \& Lodders, in prep.).  

We predict all cloud properties using the model of \citet{AM01}, which is directly coupled to the radiative transfer solver.  The model parameterizes the efficiency of sedimentation of cloud particles through an efficiency factor, $f_{\mathrm{sed}}$.  Large values of $f_{\mathrm{sed}}$ correspond to rapid particle growth and large mean particle sizes.  In this case condensates quickly settle out of the atmosphere, leading to physically and optically thin clouds.  When $f_{\mathrm{sed}}$ is small, cloud particles grow more slowly and the atmospheric condensate load is larger and the clouds are both physically and optically thicker.  Here we use a sedimentation efficiency parameter $f_{\mathrm{sed}}=3$,  which fits spectral observations of cloudy L-dwarfs \citep{Marley02}.  This choice generally confines 90\% of the optical depth of a cloud within 1 scale height of the cloud base.  While a large number of condensates are included in the calculation of the chemical equilibrium, the cloud model here accounts for the major condensates Fe and Mg-silicates (here represented by Mg$_2$SiO$_4$). Further details on our methods can be found in \citet{Marley02} and Marley et al.~(in prep).

We model the impinging stellar flux from 0.26 to 10.0 $\mu$m and the emitted thermal flux from 0.26 to 325 $\mu$m.  All the relevant planetary and stellar parameters for \hh~are taken from \citet{Sato05}.  The incident stellar flux from HD 149026 is from a model spectrum by \citet{Kurucz93} with $T_{\rm eff}=6200\,$K model interpolated in gravity to $\log g=4.26$.  The luminosity of the final stellar model matches the \citet{Sato05} derivation of $2.7\,L_\odot$.  The planet's radiative-convective \emph{P--T} profile is arrived at iteratively until the net flux is conserved to at least one part in $10^6$ in each layer.  Our model atmosphere code computes the \emph{P--T} profile and a low resolution spectrum.  A high-resolution spectrum is generated from the \emph{P--T} profile and a full line-by-line radiative transfer code, using the same chemistry and opacity databases and converged cloud properties.

When modeling the atmospheres of irradiated objects with a one-dimensional, plane parallel atmosphere code it is necessary to weigh the stellar flux $F$ by a geometric factor given by the ratio of the cross-section area of the planet to the surface area of the emitting planetary surface.  If the planet is able to reradiate the absorbed stellar energy over its entire surface (4$\pi$ steradians), an incident stellar flux of $F/4$ is used.  This assumes a very efficient horizontal redistribution of the absorbed stellar energy over the entire atmosphere.  In the other limit, redistribution is poor and the planet is only able to reradiate this flux on the day side (2$\pi$ steradians) and the normalization in $F/2$.  Here our standard case is $F/4$.  All models shown will use $F/4$ unless they are labeled ``2$\pi$,'' in which case $F/2$ is used.  Models of the atmospheres of \T~and \hd~are in better agreement with the \emph{Spitzer} observations \citep{Charb05,Deming05b} when efficient redistribution ($F/4$) is assumed \citep{Fortney05,Seager05,Barman05}.  This conclusion is tentative however, and \citet{Burrows05b} argue for a value between $F/4$ and $F/2$.  An intermediate value (although closer to $F/4$) is supported by the most recent dynamical models for \hd~\citep{Cooper05} that predict a temperature contrast between the ``hot'' and ``cold'' sides of the planet of $\sim$500 K, which may be blown by supersonic winds 60$^\circ$ of longitude from the substellar and anti-substellar points.

These dynamical models \citep[e.g.][]{Cooper05,Showman02} and the work of \citet{Iro05} and \citet{Seager05} show these flux redistribution issues are likely to be more complex that than a simple $F/4$ or $F/2$ factor.  At low pressures (less than tens of millibar) the radiative time constant is short compared to likely mixing timescales, and the atmosphere should quickly adjust to the local stellar irradiation, meaning the hottest point will be the planet's subsolar point.  At higher pressures, where the atmosphere is optically thick, the radiative time constant scales roughly as the pressure squared \citep{Iro05}.  Consequently, at these larger pressures, even very slow circulation would homogenize the temperatures around the planet.  At intermediate pressures, the interplay between circulation and radiation could lead to strong temperature inhomogeneities around the planet.  This could lead to a \emph{P--T} profile that is very similar to the $F/4$ solution at depth, but is quite different at lower pressures.  The potential exists for some regions of the day side to have a hot stratosphere while others do not.  

\subsection{Profiles, Composition, \& Energy Balance}
\begin{figure}
\plotone{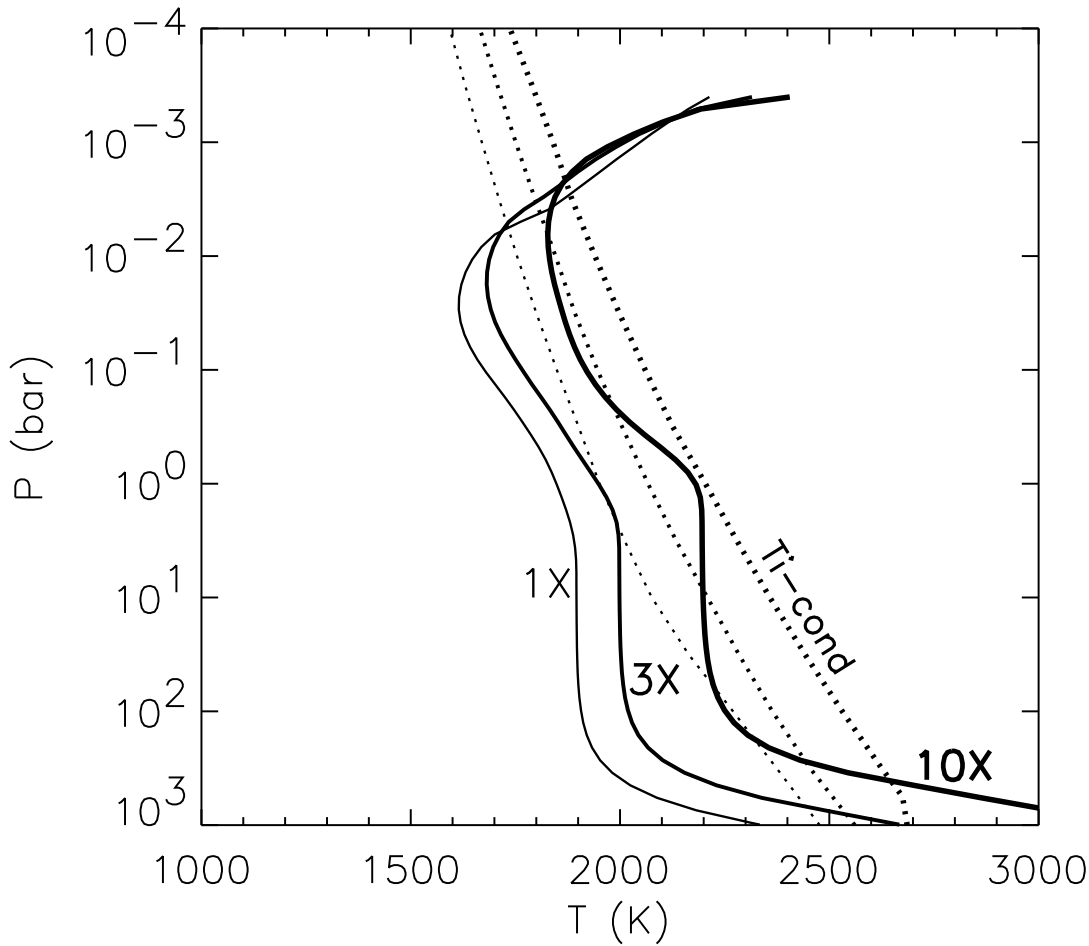}
\caption{Cloud-free atmospheric \emph{P--T} profiles for \hh~at [M/H]=0, 0.5, and 1 (1X, 3X, and 10X, respectively).  The thicker the curve, the higher the metallicity.  These profiles include the opacity of TiO and VO at low pressure, based on our pre-tabulated abundance tables, which is probably not realistic for these profiles. (See text.)  Metallicity-dependent condensation curves for the first condensate of Ti are shown as dotted lines and labeled ``Ti--cond''  Again, the thicker the curve, the higher the metallicity.
\label{figure:pt1}}
\end{figure}

We have computed atmospheric \emph{P--T} profiles for three metallicities: [M/H]=0, 0.5, and 1.  These profiles are shown in \mbox{Figure~\ref{figure:pt1}}.  The solar metallicity model allows a comparison with published models of other planets, the [M/H]=0.5 is close to the metallicity of the star HD 149026, and [M/H]=1 represents an enrichment of $\sim$4 times over the host star's value, which is typical of the atmospheres of Jupiter and Saturn.  These atmosphere models assume 1) reradiation of the absorbed stellar flux over 4$\pi$ steradians, 2) cloudless structures, and 3) an intrinsic temperature (the $T_{\mathrm{eff}}$ the planet would have in isolation) of 100 K, which we will later show to be reasonable, based on thermal evolution models (\S3.1).  As the metallicity is increased the atmosphere becomes warmer as a larger fraction of incident radiation is absorbed.

These profiles all exhibit hot stratospheres due to absorption by TiO and VO at mbar pressures.  In fact, however, the \emph{P--T} profiles all cross the condensation curve of Ti-bearing condensates at pressures of hundreds of bars.  As discussed in \citet{Lodders02b}, the first Ti condensate will not necessarily be CaTiO$_3$; the condensation sequence is far more complicated.  For solar metallicity the first condensate is TiN if $P \gtrsim 30$ bars, Ca$_3$Ti$_2$O$_7$ if $5 \lesssim P \lesssim 30$ bars, Ca$_4$Ti$_3$O$_{10}$ if $0.03 \lesssim P \lesssim 5$ bars, and CaTiO$_3$ if $P \lesssim 0.03$ bars.  These four condensates are the \emph{initial} condensates as function of total pressure and their condensation temperatures define the Ti-condensation curve in \mbox{Figure~\ref{figure:pt1}}.  At higher metallicities the situation is similarly complex.

Due to mixing in the atmosphere, any gaseous TiO in the upper atmosphere is expected to eventually mix down to this condensation point, where the Ti-bearing cloud remains confined due to the planet's gravitation field.  This process, known as a ``cold trap,'' is responsible for the extremely low water abundance in the Earth's stratosphere.  The abundance of TiO at pressures above the cloud would be set by the vapor--solid chemical equilibria between the condensate and the TiO gas.  Below we will discuss problems with calculating this equilibrium for all possible \emph{P--T} profiles.  The profiles shown in Figure~\ref{figure:pt1} are potentially unrealistic as the actual TiO abundances high in the atmosphere should be less than what we have used (see below).

The formation of a hot stratosphere arising from TiO and VO absorption at low pressures and high temperatures was pointed out for highly irradiated planetary atmospheres by \citet{Hubeny03}, and we obtain qualitatively similar results.  A full discussion of titanium and vanadium chemistry in the context of low mass dwarf stellar atmospheres can be found in \citet{Lodders02b}.  Much of that discussion pertains to the atmospheres of Pegasi planets as well, but here we encounter additional issues. The cold trap phenomenon constitutes a departure from chemical equilibrium that cannot be accounted for in pre-tabulated chemical equilibrium abundances \citep{Hubeny03}.  Our chemical abundances and opacities are pretabuled in \emph{P--T} space and the atmosphere code interpolates in these abundances as it converges to a solution.  The abundances determined for any one pressure level of the \emph{P--T} profile are not cognizant of abundances of other levels of the \emph{P--T} profile.  In this case a tabulated TiO abundance at a given \emph{P--T} point at which, in equilibrium, TiO would be in the gaseous state, may not be correct given the behavior of the profile at higher total pressures, where the profile may cross a condensation curve that would deplete the atmosphere of TiO above the cloud.


As described in \citet{Lodders02b} gaseous VO (whose abundance is $\sim$10\% that of TiO) does not condense as solid VO, but instead condenses with TiO into the solid Ca-Ti-oxides.  This is because V substitutes for Ti in these condensates, which condense at higher temperatures than solid VO.  It is thus reasonable to use the Ca-Ti-oxides condensation curve to determine the pressures and temperatures at which TiO and VO begin to condense.  We stress that in our chemical equilibrium calculations the metallicity dependence of condensation is handled appropriately for every metallicity.  We suggest the reader turn to Figures 3 and 4 in \citet{Lodders02b} which show contours of constant TiO and VO mole fraction at a wide range of temperatures and pressures.  For instance, in the solar metallicity case, the TiO mole fraction at 100 mbar drops by a factor of 100 between 1850 K, when Ca$_4$Ti$_3$O$_{10}$ condensate forms, and 1650 K, a point on our \emph{P--T} profile.

Given the difficulty of treating the TiO/VO condensation issue and because this is the first study of the atmosphere of \hh, we have decided to take a simpler path.  Here for most of our profiles we will bracket the effect that gaseous TiO and VO would have on the atmosphere.  We do this by choosing two endmember cases for the TiO/VO abundances.  We either leave TiO and VO in the opacity table, as dictated by the chemistry table, or we remove all TiO and VO opacity at pressures less than 10 bars.  (Essentially all stellar flux is absorbed at $P < 10$ bars.)  A somewhat similar approach was taken by \citet{Hubeny03} in their models of the atmosphere of planet OGLE-TR-56b.  These different assumptions give profiles that either have hot stratospheres, as shown in \mbox{Figure~\ref{figure:pt1}}, or lack these stratospheres, as discussed below.  These two limiting cases likely bracket the range of possible profiles for \hh.  In one case we will use a more complex approach.  As seen in \mbox{Figure~\ref{figure:pt1}} the [M/H]=1.0 profile reaches the [M/H]=0.0 condensation curve.  This means that while $\sim$90\% of TiO condenses, 10\% of the TiO (a solar abundance) will be left in the atmosphere.  Therefore we will also compute a profile that uses opacities based on the following abundances:  [M/H]=1.0 is used for all species except TiO and VO, where abundances from the [M/H]=0.0 table are used at $P$<10 bars.

\begin{figure}
\plotone{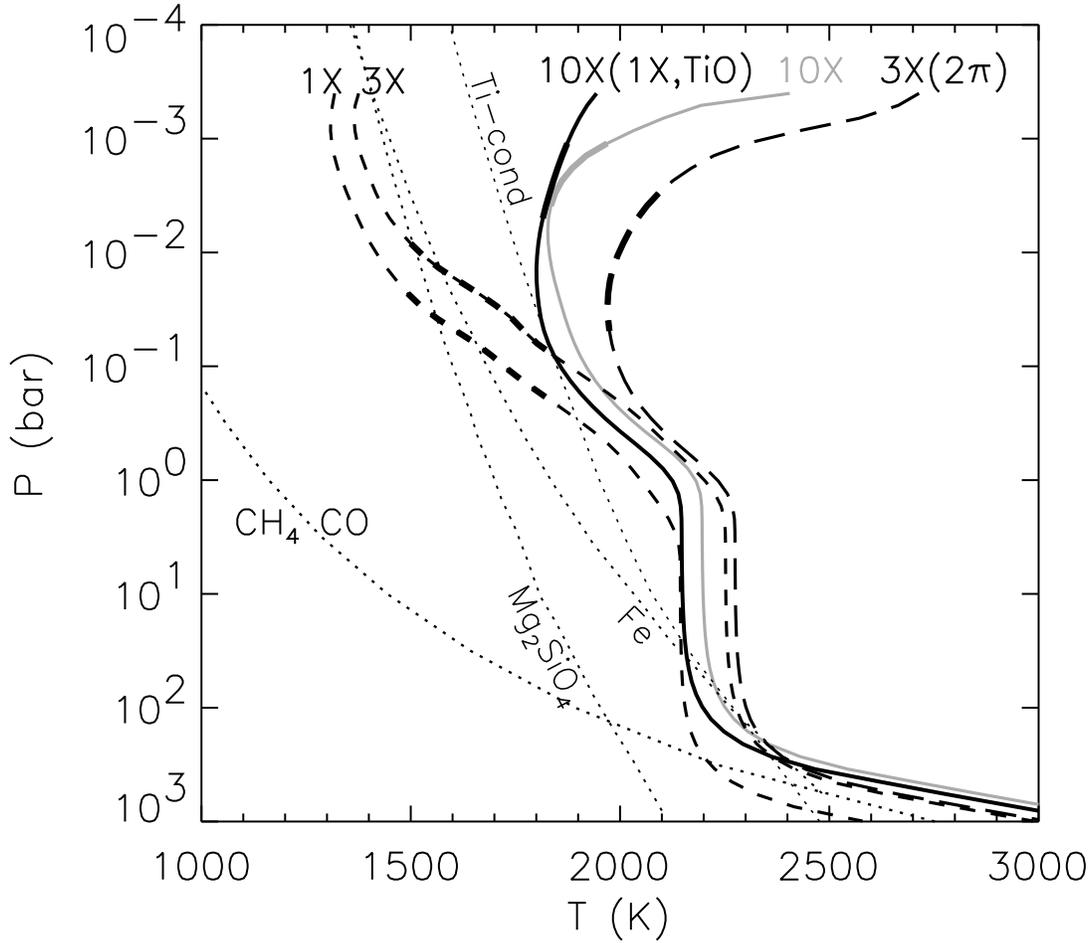}
\caption{Atmospheric \emph{P--T} profiles for \hh. In the models with [M/H]=0 (``1X'') and 0.5 (``3X''), the opacity of TiO and VO has been removed at pressures less than 10 bar, but the opacity of clouds of forsterite and iron is included.  The [M/H]=1 model in grey is the same as that shown in \mbox{Figure~\ref{figure:pt1}}.  The ``3X(2$\pi$)'' model assumes that the absorbed stellar flux can only be reradiated on the planet's day side and that the chemical equilibrium TiO and VO abundances are valid at low pressure.  The profile labeled ``10X(1XTiO)'' is uses [M/H]=1.0 table for all abundances for except for TiO and VO, for which the [M/H]=0 abundances are used.  The thick part of each profile shows the extent of the brightness temperatures for each model in the 3 to 30 $\mu$m wavelength range.  Also shown is the curve where the abundances of CH$_4$ and CO are equal.  For clarity, all condensation and chemical boundaries are for \emph{solar composition} \citep{Lodders05}.   
\label{figure:pt2}}
\end{figure}

Two of the modified profiles lacking TiO and VO at low pressure are shown in \mbox{Figure~\ref{figure:pt2}}.  They are for [M/H]=0.0 and 0.5 and include the opacity of Mg-silicates (here as forsterite, Mg$_2$SiO$_4$) and iron (Fe) clouds.  These cloudy profiles are slightly warmer at low pressures than cloudless profiles would be.  Both models absorb nearly all incident stellar flux, with only 8\% and 5\% of this flux being scattered back to space, for [M/H]=0 and 0.5, respectively.\footnote{These values are not necessarily equivalent to the Bond albedo, as we are using a plane parallel, one-dimensional model atmosphere code.}  The stellar flux is absorbed at progressively lower pressures as the metallicity is increased.  We also show a [M/H]=0.5 model assuming reradiation over the day side only (2$\pi$ steradians).  In this model, we also leave in our pre-tabulated TiO and VO abundances, to create what may be a reasonable upper limit to temperatures on the day side, for a uniform [M/H]=0.5 atmosphere.  We also include the [M/H]=1.0 model from \mbox{Figure~\ref{figure:pt1}} here, shown in grey.  We compare that profile to the [M/H]=1.0 profile with solar TiO and VO abundances in the upper atmosphere, shown in solid black.  With the smaller TiO/VO abundances, the stratosphere persists but is significantly muted.  We will treat these profiles shown in \mbox{Figure~\ref{figure:pt2}} as respresentative of the range of potential day-side \emph{P--T} profile, while knowing that the actual temperature structure may be more complicated.  However, since it is unlikely that the planet has a lower metallicity than the parent star ([Fe/H]=$0.36\pm0.05$, \citealt{Sato05}), for the remainder of the discussion we will consider only the [M/H]=0.5 and 1 models.

A cold trap in this \emph{P--T} space will in general affect additional chemical species such as FeH and CrH, but this is of minor importance for our profiles.  However, cold traps do have the potential to change the abundance of oxygen in the upper atmosphere.  One could envision a cooler profile where silicates condensed at high pressure (here they do not), but then at lower pressures the temperature profiles could be warmer than the silicate condensation curves.  If so, then our chemical equilibrium abundance tables would significantly overestimate the abundance of available oxygen.  The Mg-silicates remove 23\% of oxygen in a solar composition gas, which would increase the C/O ratio in the upper atmosphere.  This would lead to a decrease in the mixing ratio of H$_2$O and an increase in the mixing ratio of CO \citep{Fortney05}.  Such concerns may be relevant to atmospheres encountered in the future.

\subsection{Spectra}
\begin{figure}
\plotone{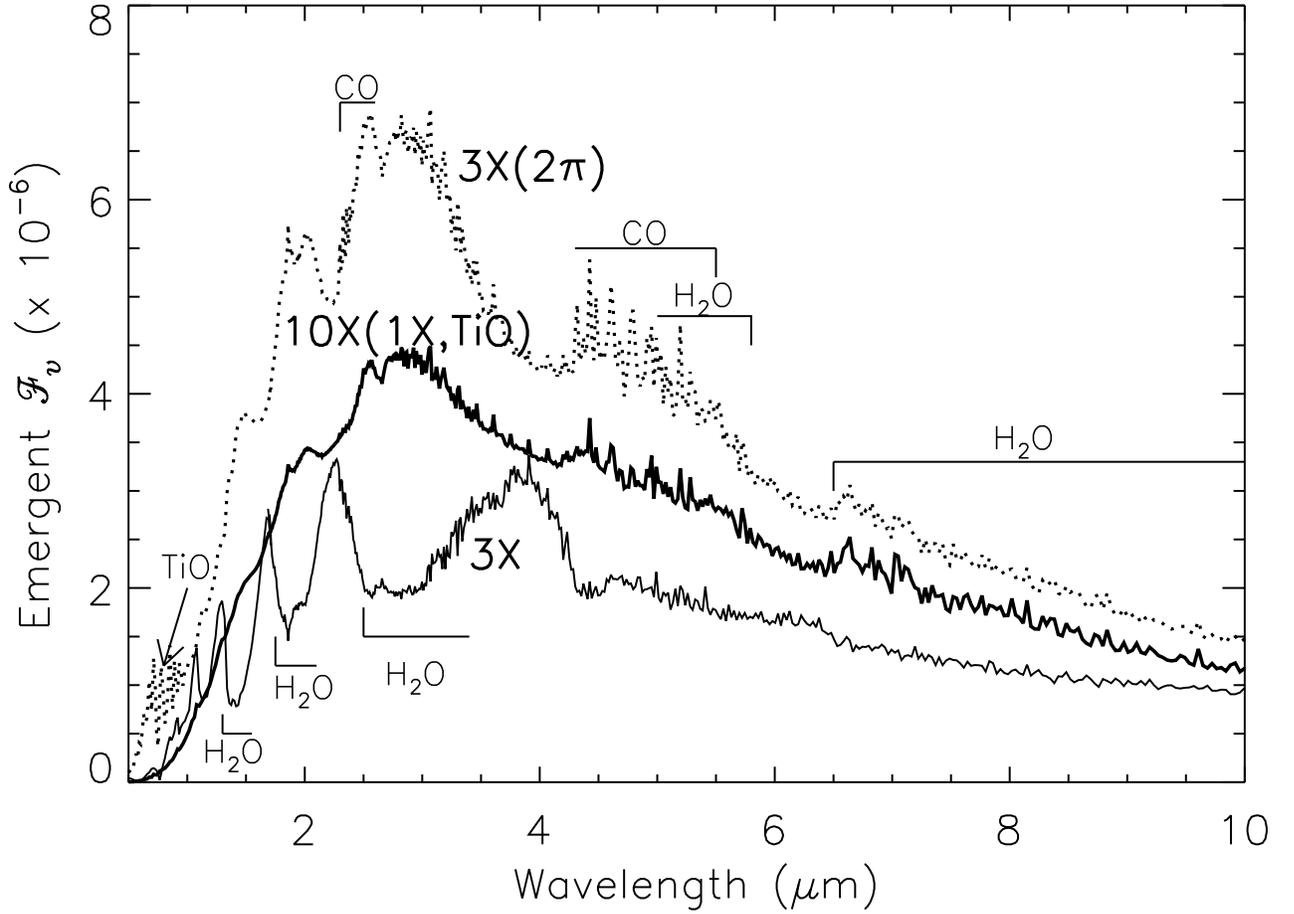}
\caption{Emergent infrared spectra for \hh~at [M/H]=0.5 (labeled 3X), [M/H]=1.0 with TiO/VO reduced to solar (labeled 10X(1X,TiO)), and [M/H]=0.5 with 2$\pi$ steradian reradiation (labeled 3X2$\pi$).  The flux is in erg s$^{-1}$ cm$^{-2}$ Hz$^{-1}$.  Many features seen in absorption in the [M/H]=0.5 model can be seen in emission in the other models.  Most prominent are the H$_2$O and CO features in the infrared.  TiO can be seen in emission in the optical in the [M/H]=0.5 2$\pi$ model, whereas the [M/H]=0.5 model has strong absorption due to Na and K in the optical.
\label{figure:spec}}
\end{figure}

\begin{figure}
\plotone{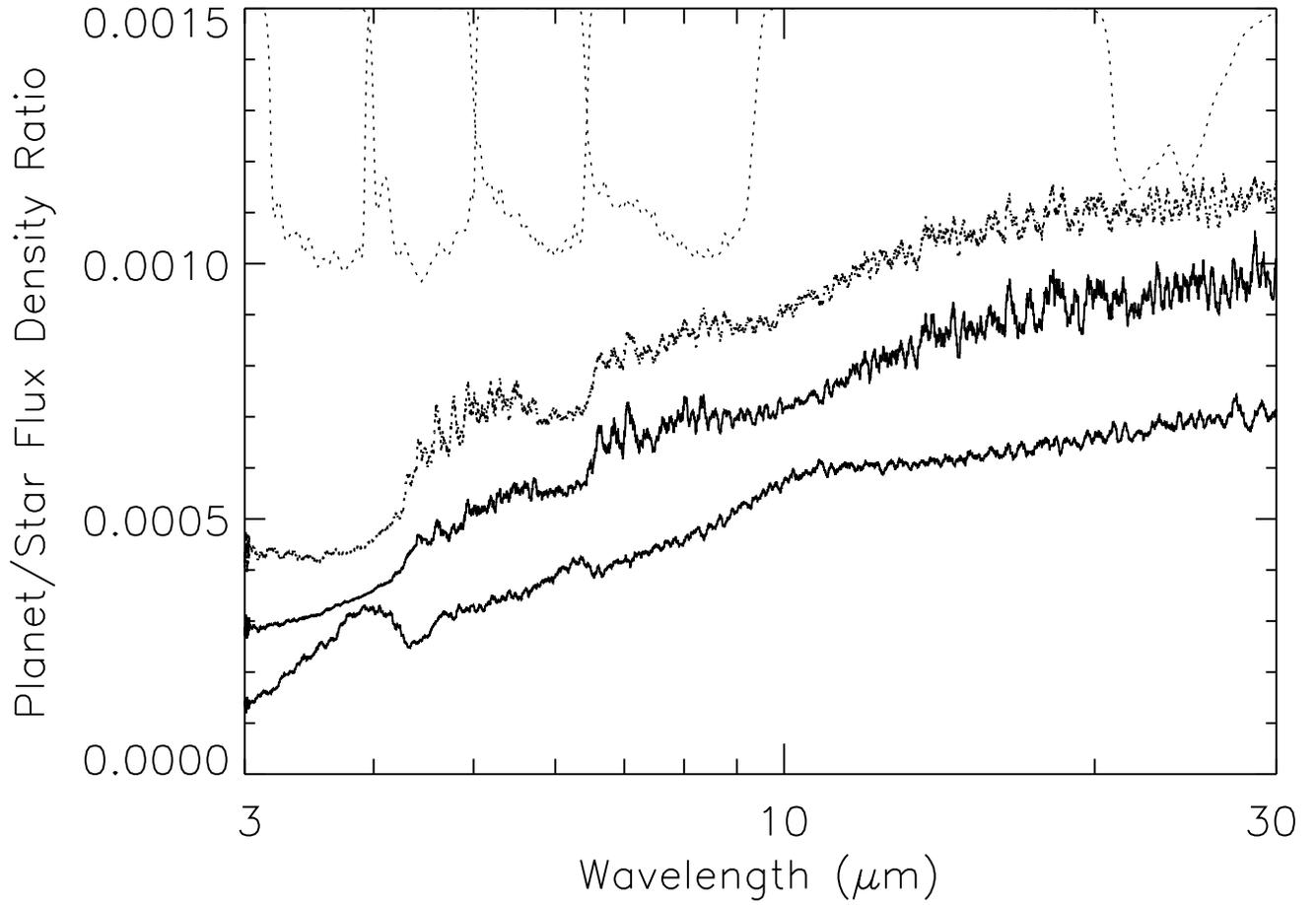}
\caption{Planet-to-star flux density ratios for the three \hh~planet models shown in \mbox{Figure~\ref{figure:spec}}.  The lower black curve is the cloudy model with [M/H]=0.5.  The upper black curve is the [M/H]=1 model with the TiO/VO reduced to solar.  The dotted curve is the model with [M/H]=0.5 and 2$\pi$ steradian reradiation.  Shown in dotted lines, hanging from above, are the sensitivity functions of the 4 IRAC bands and the MIPS 24 $\mu$m band.
\label{figure:rat}}
\end{figure}

The spectra of the [M/H]=0.5 and 1 models are plotted in \mbox{Figure~\ref{figure:spec}}.  Like other Pegasi planets, the infrared spectrum is carved by H$_2$O and CO opacity.  We find no spectral signatures due to CH$_4$ as CO is by far the dominant carbon reservoir in these hot atmospheres.  Interestingly, models with hot stratospheres result in \emph{emission} features of H$_2$O and CO.  For example, the series of spikes in the 4.2 - 5.5$\,\mu$m spectral region are all CO emission features.  Infrared emission features have long been known in the atmospheres of our solar system's giant planets.  All four planets possess a hot stratosphere and show a CH$_4$ emission feature at the molecule's strong 7.8 $\mu$m band.  Photochemically produced ethane and acetylene also exhibit emission in some giant planet atmospheres.  For CH$_4$, only the 7.8 $\mu$m band is seen in emission because it is the strongest band, and it forms highest in atmosphere---in this case, above the temperature inversion.  For \hh~models with the hottest stratospheres, emission due to TiO and VO is also seen in the optical, whereas in the [M/H]=0.5 model the optical spectrum is carved by strong absorption by Na and K.  We plot our planet-to-star flux density ratios in \mbox{Figure~\ref{figure:rat}} and show the wavelength ranges of the 4 IRAC bands and MIPS 24 $\mu$m band.  The models with stratospheres are redder from 4 to 20 $\mu$m than models without stratospheres, due to emission by CO and H$_2$O, which can be seen most strongly in IRAC 4.5, 5.8, and 8.0 $\mu$m bands.  Comparing photometry in these IRAC bands to the 3.6 $\mu$m band will be the most clear signature of a stratosphere for this planet, as well as other transiting planets.

Table 1 collects our band-averaged planet-to-star flux density ratios in 
the \emph{Spitzer} IRAC and MIPS bands.  Due to the combination of the 
planet's small radius and the large luminosity of the parent star, these 
ratios are typically $\sim$1/4 to 1/2 of those for \T, assuming 
``4$\pi$'' reradiation.  Detection may be difficult.  However, we will show in \S2.5 that ratios from 3 to 5 $\mu$m for models with hot stratospheres are comparable to \T.  Deciding the best 
bands to search in will depend on the band-dependant photon-noise 
levels.\footnote{Discussions of \emph{Spitzer} IRAC and MIPS sensitivities can be found at http://ssc.spitzer.caltech.edu/irac/sens.html and http://ssc.spitzer.caltech.edu/mips/sens.html, respectively.}   Looking at IRAC specifically, while the planet is 
brightest at 8.0 $\mu$m, the instrument is more sensitive in the shorter 
wavelength bands.  As \T~seems to be redder than expected in the infrared 
\citep{Fortney05,Barman05} we look forward to the comparison of models with the photometry in all four IRAC bands as that data becomes available.

\subsection{Limb Darkening and Brightening}
\begin{figure}
\plotone{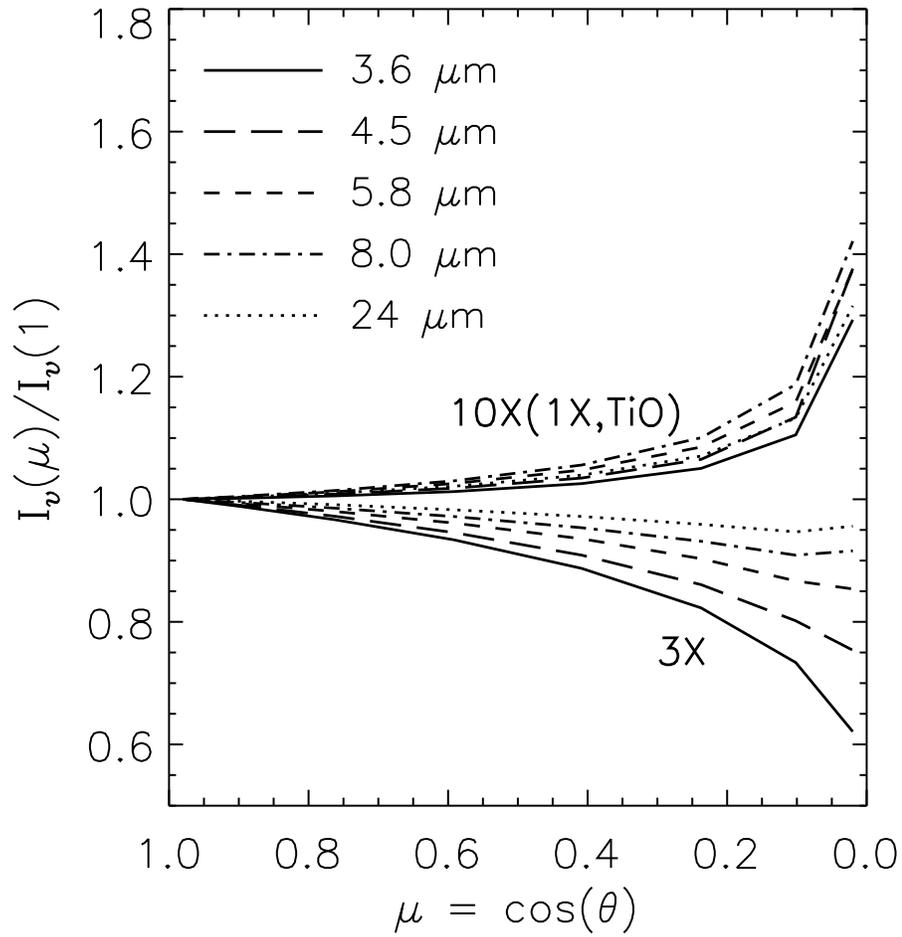}
\caption{Limb darkening profile for two atmosphere models of \hh~in 5 different \emph{Spitzer} bandpasses.  The ``10X(1X,TiO)'' curves are for the [M/H]=1 atmosphere models (with TiO and VO solar) with a hot stratosphere.  The ``3X'' curves are for the cloudy, [M/H]=0.5 atmosphere models.  (See \mbox{Figure~\ref{figure:pt2}}.)
\label{figure:limbs}}
\end{figure}

A diagnostic of the atmospheres of transiting planets is the variation in surface brightness of the planet's disk which is in principle observable during the secondary eclipse ingress and egress.  Limb darkening is particularly interesting as it provides an empirical determination of the temperature profile of the atmosphere at optical depths less than unity.  We have computed the planet's limb darkening over the \emph{Spitzer} bandpasses.  Depending on the atmosphere model, both limb brightening and darkening can occur.  Particularly, we find that limb brightening is a sensitive diagnostic of stratospheric heating, with a strong signature in the [M/H]=1 model.  \mbox{Figure~\ref{figure:limbs}} plots the normalized specific intensity across the planetary disk for our cloudy [M/H]=0.5 and the cloudless [M/H]=1 models.  We have used the transit code described in \citet{Hubbard01} to compute ingress lightcurves for the limb brightened and darkened planets.  Unfortunately the difference between these lightcurves is only $\sim$1\%, so the effect is not detectable with \emph{Spitzer} but probably within reach of the \emph{James Webb Space Telescope}.  Therefore, for the foreseeable future, IRAC photometry remains the best method to search for stratospheres in Pegasi planets..

\subsection{Comparative Atmospheres: HD 189733b}
Much of our understanding of Pegasi planet atmospheres will come from comparing and contrasting atmospheres within the class that may differ in the amount of stellar flux received or in atmospheric abundances.  Here we briefly compare the atmosphere of \hh, relative to \hd, \T, and \he, which was recently discovered by \citet{Bouchy05}.

\begin{figure}
\plotone{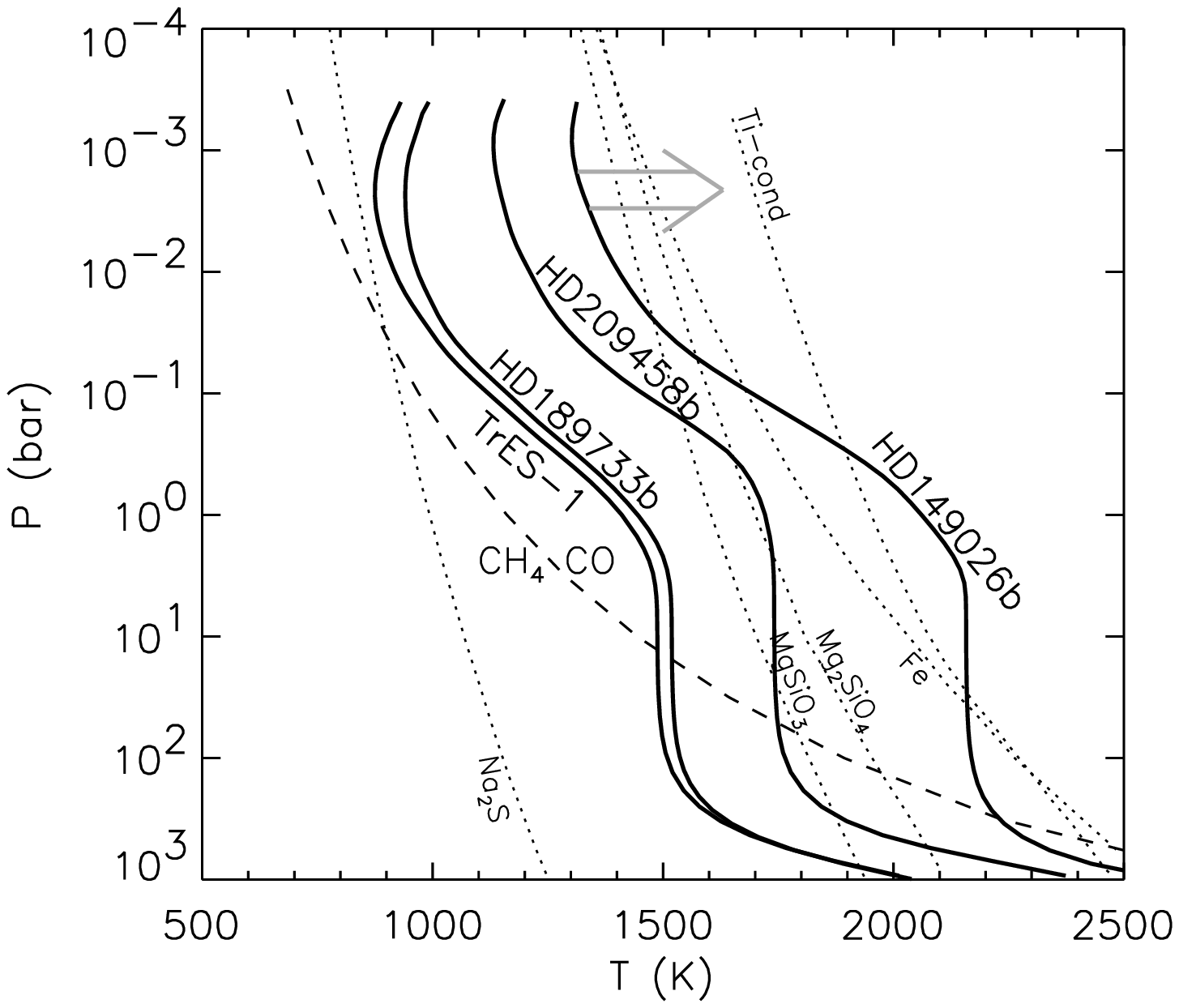}
\caption{Atmospheric \emph{P-T} profiles for planets \T, \he. \hd, and \hh, from left to right.  All profiles assume solar metallicity and reradiation of absorbed stellar flux uniformly over 4$\pi$ steradians.  Condensations curves are labeled and are shown as dotted curves and the boundary where CO and CH$_4$ have the same abundance is the dashed curve.  The grey arrow attached to the \hh~profile signifies that this is likely a lower limit to the stratospheric temperatures.  All profiles assume $T_{\rm int}$=100 K.
\label{figure:ptall}}
\end{figure}

Planet \he~orbits an early main sequence K star at 0.0313 AU.  Although dubbed a ``very hot Jupiter'' due to its short orbital period of 2.219 days, because of the small stellar luminosity ($\sim$0.34 L$_{\odot}$) we find that its atmosphere is colder than \hd~and only marginally warmer than that of \T.  A comparison of our derived \emph{P-T} profiles for \he~and \hh~with other Pegasi planets is shown in \mbox{Figure~\ref{figure:ptall}}.  For simplicity and for comparison purposes, these profiles all assume solar metallicity and reradiation over 4$\pi$ steradians.  As with \T, it appears that Mg-silicate and iron clouds form too deep in the atmosphere of \he~to affect its spectrum.

\begin{figure}
\plotone{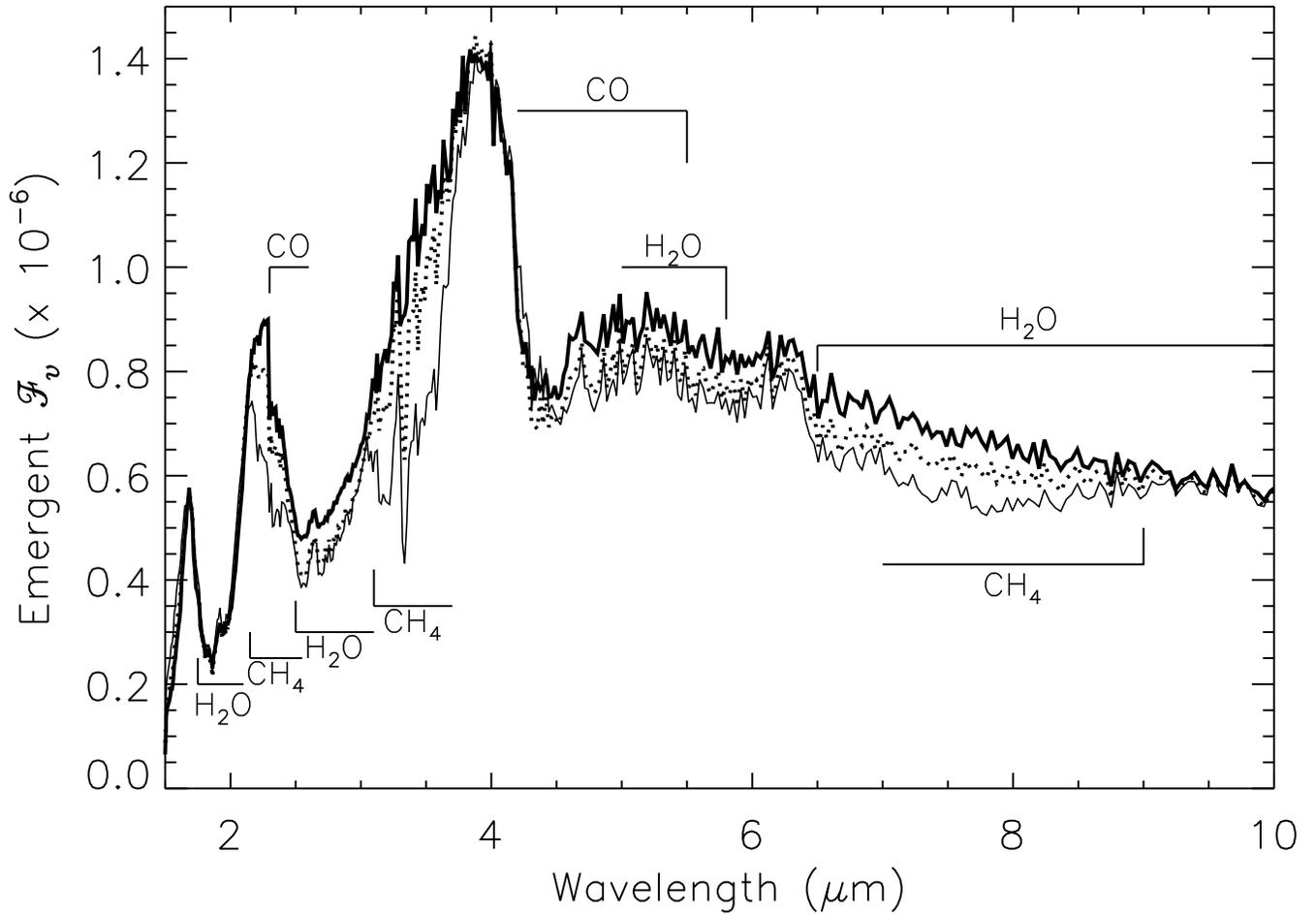}
\caption{Emergent spectra of planet \he~at three metallicities.  The thin solid line is for [M/H]=0.0.  The dotted line is for [M/H]=0.5.  The thick line is for [M/H]=1.0.  Relevant absorption features are labeled.  Absorption due to CH$_4$ is very sensitive to the atmospheric metallicity, especially from 3 to 4 $\mu$m.
\label{figure:spec189}}
\end{figure}

Similar to our findings for \T~in \ct{Fortney05}, we find that CH$_4$ will be abundant in the atmosphere of \he, although CO is still be the major carbon carrier.  As discussed in \S2.2 and \ct{Fortney05}, condensation curves and chemical boundaries are functions of the atmospheric metallicity.  The specifics of CO and CH$_4$ abundances are discussed in detail in \ct{Lodders02}.  If the atmosphere of \he~is enhanced in heavy elements, then the atmospheric abundance of CH$_4$ will be significantly reduced, for the two following reasons.  At higher metallicity, atmospheric opacity is higher, leading to greater absorption of stellar flux, and a warmer atmosphere.  This favors CO over CH$_4$.  In addition, the curve where CO and CH$_4$ have the same abundance moves to the left in \mbox{Figure~\ref{figure:ptall}} as metallicity increases, towards lower temperatures at a given pressure.  This again leads to a larger CO/CH$_4$ ratio.  Dramatic effects on the infrared spectra can be seen in \mbox{Figure~\ref{figure:spec189}}.  While methane absorption is strong in bands centered at 2.2, 3.3, and 7.8 $\mu$m in the [M/H]=0.0 model, this absorption fades significantly at [M/H]=0.5 and [M/H]=1.0.

A low level of flux in the 3.6 $\mu$m IRAC band for \he~or for \T, due to methane absorption, would have significant implications for these planet's atmospheres.  First, it would hint that these atmospheres do not have a significantly supersolar metallicity.  Second, it would support our derived \emph{P-T} profiles that assume extremely efficient redistribution of absorbed stellar flux onto the night side.  In the limit that this redistribution only occurs over the illuminated hemisphere, the day side would be $\sim$200-250 K hotter (see Figure 1 in \citealt{Fortney05} for \T) and therefore only spectroscopically insignificant amounts of methane would be present.  A detection of methane would seem to indicate that the atmosphere is efficient at dynamically redistributing energy onto the night side.  A more detailed calculation would include the kinetics of the CO/CH$_4$ reaction.  Work along these lines is currently underway for planet \hd~which indicates that the CO/CH$_4$ is enhanced over equilibrium values and is nearly constant around the entire planet \citep{Cooper05b}.

\begin{figure}
\plotone{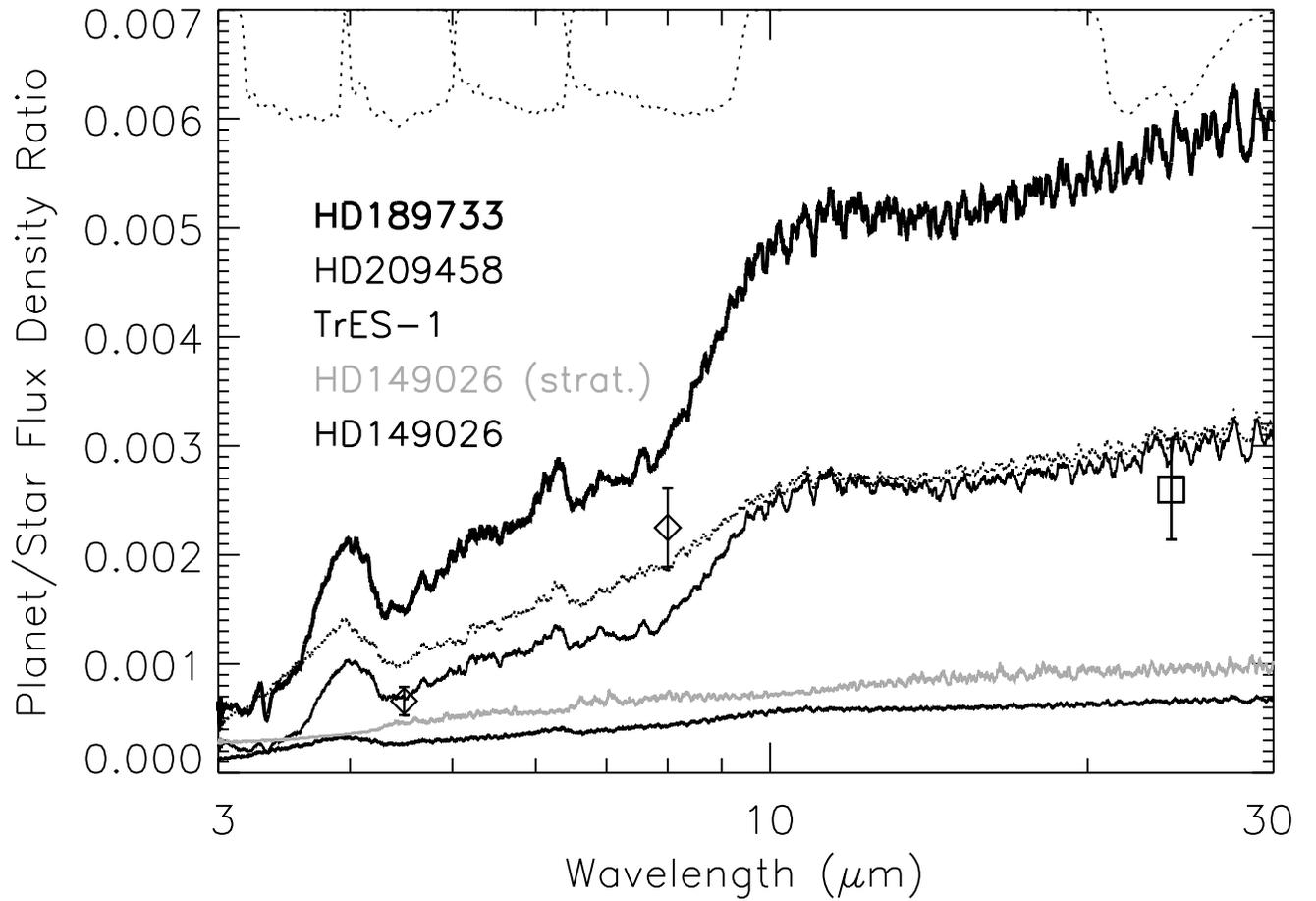}
\caption{Planet-to-star flux density ratios for the four known transiting planets around bright stars.  From top to bottom, HD 189733, HD 209458, \T, HD 149026 with a stratosphere (in grey), and HD 149026 without a stratosphere.  All atmospheres assume [M/H]=0.0 for simplicity, except the \hh~models, for which [M/H]=1.0 (but solar TiO/VO), in grey, and [M/H]=0.5.  Data at 4.5 $\mu$m and 8 $\mu$m (diamonds) are for \T~and datum at 24 $\mu$m (square) is for HD 209458.  1$\sigma$ error bars are shown.
\label{figure:compare}}
\end{figure}

Interestingly, the planet-to-star flux density ratios in the HD 189733 system should be much larger than in any other bright transiting system, since the stellar radius is only $0.76 \pm 0.01$ R$_{\odot}$ and the planetary radius is $1.26 \pm 0.03$ \rj.  The ratio of the cross-sectional areas (the measured quantity) is 2.9\% \citep{Bouchy05}.  In \mbox{Figure~\ref{figure:compare}} we compare planet-to-star flux density ratios for HD 189733, HD 209458, \T, and HD 149026.  In general the ratios for the HD 189733 system are twice those for HD 209458 and \T.  In Table 1 we collect our band-integrated planet-to-star flux density ratios in the \emph{Spitzer} IRAC and MIPS bands.  From 3 to 3.8 $\mu$m, and around 7.8 $\mu$m \he~shows absorption due to CH$_4$, which is slightly weaker in strength than in \T, due to the 50 K higher $T_{\rm eff}$.  From 3 to 3.8 $\mu$m, this band appears closer to saturation in \T, as a wider absorption feature is seen.  Strong CO absorption is seen near 4.5 $\mu$m, which is expected for all Pegasi planets.  Much of the rest of the spectrum is carved by H$_2$O absorption as shown in \mbox{Figure~\ref{figure:spec}} and \mbox{Figure~\ref{figure:spec189}}.  For \hh, we can see that models with hot stratospheres may appear nearly as bright as \T~in the 3.6 $\mu$m IRAC band.

\section{Evolution Models}
\subsection{Methods}
We have generated a grid of atmospheres for the purpose of modeling the evolution of \hh.  This is a grid of cloudless atmospheres with [M/H]=0.5 and TiO and VO removed at $P$<10 bar.  The stellar flux is diluted by a factor of 1/4, as described in \S2.1.  As we predict that the silicate and iron clouds have a combined optical depth of $\sim$0.1, neglecting the clouds leads to a very slightly cooler \emph{P-T} profile in the upper atmosphere but the structure of the deep atmosphere is essentially unchanged from the profile shown in \mbox{Figure~\ref{figure:pt2}}.  We assume that the stellar luminosity is constant with time and that the planet has remained at 0.042 AU for all times.  We have calculated atmosphere models at 12 values of $T_{\mathrm{int}}$ and 3 gravities, that cover the parameters of planet \hh~for ages between $\sim$10$^6$ to $5\times10^9$ years.  Even at very young ages ($<10\,$Myr), when the planet has an appreciable internal energy source, the incident flux dominates the planet's intrinsic flux, and $T_{\mathrm{eff}}$ is only 30 K higher than $T_{\mathrm{eq}}$.

Our evolution code for the calculation of the cooling and contraction of adiabatic giant planets is well-tested.  It has been used to produce evolutionary models of Jupiter and Saturn \citep{FH03} and cool extrasolar giant planets \citep{FH04} and it is described in detail in \citet{FH03} and \ct{Fortney04}.  No modifications to the code were needed to calculate the evolution and contraction of Pegasi planets. The radiative atmosphere that serves as the bottleneck for cooling above the adiabatic H/He envelope is automatically accounted for with our fully non-grey, self-consistent model atmosphere grid.  The importance of using detailed atmosphere models for evolutionary calculations of Pegasi planets is discussed in \citet{Baraffe03}.  The heat extracted from the planet's interior per unit mass is given by:
\begin{equation}
\label{S}
\frac{\partial L}{\partial m} = -T\frac{\partial S}{\partial t},
\end{equation}
where $L$ is the planet's intrinsic luminosity, $T$ is the temperature of a mass shell, $S$ is the specific entropy of a mass shell, and $t$ is the time.  In practice, a set of structural models of decreasing internal heat content determine $T$ and $S$, and the model atmosphere grid allows for the derivation of $L$ given $S$ from a structural model.  One can then solve for the time step $t$.

For the equation of state (EOS) of the H/He envelope we use the \citet{SCVH} EOS with a helium mass fraction $Y$=0.27.  In order to include heavy elements accurately we use the ANEOS zero-temperature EOSs \citep{ANEOS} of water (for generic ices, which is predominantly water) and olivine (for generic ``rocks'').  Zero-temperature EOSs are suitable for these constituents, as the thermal contribution to the pressure for ices in our density/temperature range is on the order of 10-20\% \citep{Hubbard80b} and is considerably lower for rock.  Later we will show this has little effect on our derived core mass.  The use of a zero-temperature EOS implies that we ignore the heat content of the core on the thermal evolution of the planet.  This is often done for evolutionary models of Jupiter and Saturn \citep{Hubbard77,Saumon92,FH03}, as the error involved is small compared to other unknowns, such as the temperature of the H/He envelope at Mbar pressures \citep{Saumon04} and the radiative properties of the planets' atmospheres.  At first glance, this may seem a gross simplification for \hh, where the heavy elements may constitute 2/3 of the planet's mass.  However, as noted in \citet{Hubbard78} in regard to the cooling of Uranus and Neptune, if the electrons in the interior of a planet are degenerate, then the specific heat at constant volume is determined by the ions only.  Therefore, the heat capacity of a species is inversely proportional to its molecular weight, meaning the heat content of the ices is $\sim$0.1 that of H, and for rocks this ratio is only a few hundredths.

Modeling the heat content of the core would require a knowledge of the mechanism of energy transport within the core.  If energy were transported by convection, the temperature gradient would be close to an adiabat but if conduction is the dominant mechanism, the core could be close to isothermal.  Furthermore, interior composition gradients can inhibit convection and result in steep super-adiabatic temperature gradients, as may be the case for Uranus and Neptune.  Adiabatic models for Uranus and Neptune overestimate the current luminosity of both planets \citep{Ubook}, supporting such a possibility.

It should be remembered that compared to Jupiter and Saturn, relatively little is understood about the interiors of Uranus and Neptune.  Our understanding of Jupiter and Saturn forms a solid basis for modeling the structure of extrasolar \emph{gas} giants but the same cannot be said about extrasolar \emph{ice} giants.  (For an in-depth discussion of the structure and cooling history of the solar system's ice giants, see the excellent reviews by \citealt{Ubook} and \citealt{Nbook}.)  Given the considerable uncertainties in modeling the interior structure of a planet for which only the mass and radius are known, we believe the EOSs and treatment we have chosen here are quite acceptable.

We also neglect the ``transit radius'' effect: The apparent radius of a transiting planet is the radius where the slant optical depth through the planet's atmosphere reaches unity.  The corresponding atmospheric pressure can vary across many orders of magnitude, depending on the wavelength \citep{Hubbard01,Fortney03}.  \citet{Burrows03} and \citet{Baraffe03} have estimated this effect to be $\sim$10\% and 5\% respectively, for \hd, compared to some reference radius, such as the radiative-convective boundary or the 1 bar level.  However, the transit radius effect is considerably smaller for \hh~because of its higher gravity (1700$\,$cm/s$^2$) and the broad photometric bandpasses use by \citet{Sato05}.  Furthermore, the reported error on the radius of \hh~is $\pm3.5$\%, larger than the expected transit radius effect.

In their discovery paper, \citet{Sato05} used planet modeling techniques described in \citet{Bodenheimer03} to model the contraction history and current structure of \hh.  They modeled the planet as a solar composition H/He envelope on top of a uniform density core of either 5.5 g$\,$cm$^{-3}$ (for ices) or 10.5 g$\,$cm$^{-3}$ (for rocks).  They also implicitly ignore the heat content of the core.  Constant density models ignore the actual high-pressure equation of state which could potentially lead to errors in the determination of the core mass.  However, as we will see, our estimates for the total heavy element abundances, using detailed non-grey atmosphere models and advanced equations of state for ice and rock essentially confirm the preliminary findings of \citet{Sato05}.

\subsection{Internal Structure and Composition}
\begin{figure}
\plotone{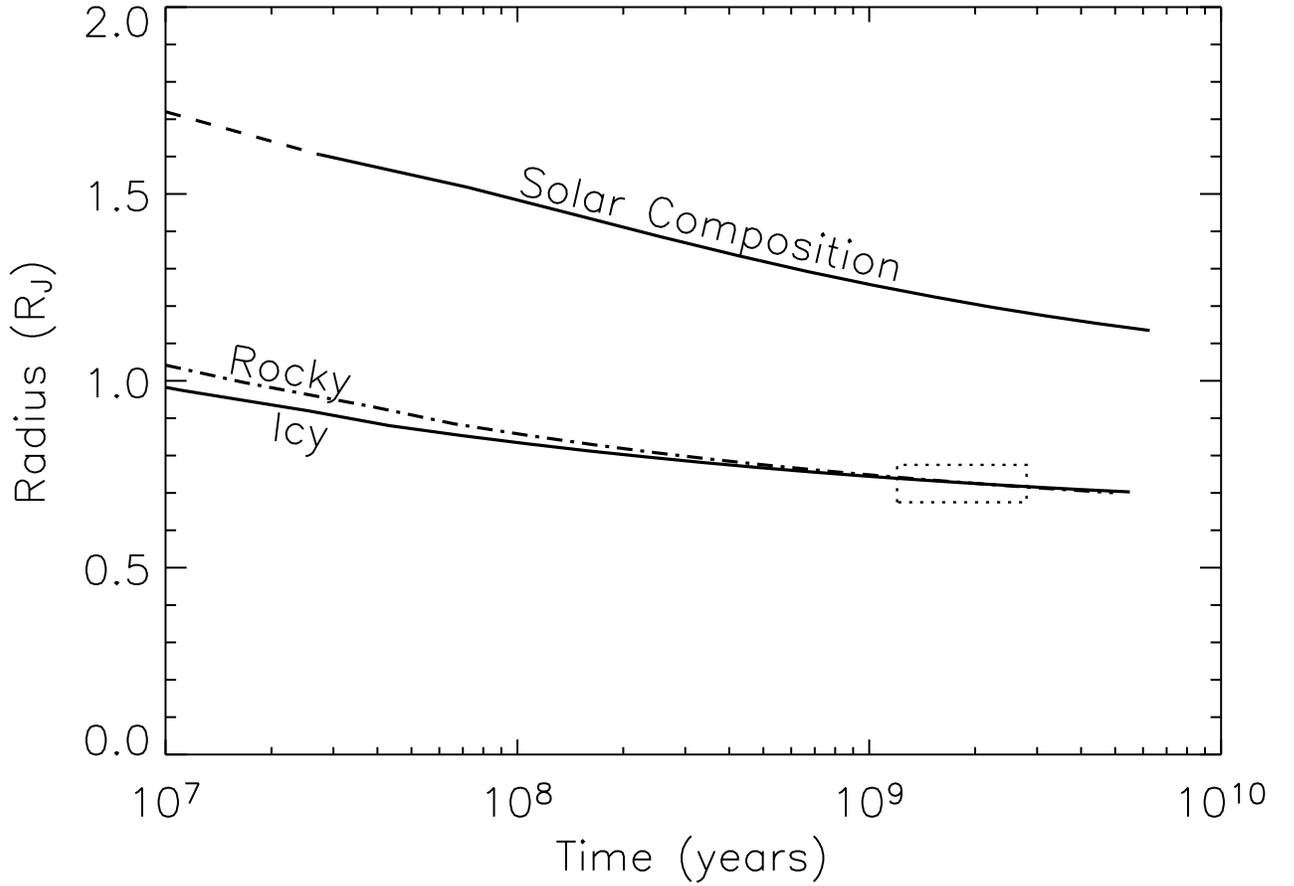}
\caption{Radii of \hh~planet models.  A planet model of solar composition is shown for comparison purposes.  The two models that match the observed radius at 2 Gyr include a H/He envelope with Z=0.045, approximately 3 times the solar metallicity, which matches the atmosphere model.  The ``Rocky'' model includes a core of 65.5 \me~of olivine, while the ``Icy'' model includes a core of 77 \me~of ice.  The age/radius error box is from \citet{Sato05}.  The dashed portion of the solar composition model indicates where the planet was off the atmosphere grid due to low gravity.
\label{figure:RvsT}}
\end{figure}

We will take a straightforward look at possible interior models of \hh.  Since we have used [M/H]=0.5 for the model atmosphere grid, we also assume the H/He envelope down to the core is similarly enhanced in heavy elements.  We use our ice EOS to mix in $Z$=0.045 into the pressure-density relation of the H/He envelope, using the additive volume rule,
\begin{equation}
\label{linear}
\frac{1}{\rho} = \frac{1-Z}{\rho_{H/He}} + \frac{Z}{\rho_{ice}}.
\end{equation}
We assume that the core is made either entirely of rock or ice.  We use a planet mass of 0.36 \mj~but vary the current age of the planet between 1.2 and 2.8 Gyr \citep{Sato05}.  Most importantly, for the planet's radius we use the 1 $\sigma$ error bars from \citet{Charb05b}, which are somewhat larger than those of \citet{Sato05}.  The largest amount of heavy elements will be needed for a young planet with a radius at the bottom of the error bar with a core composed of ice.  The smallest amount of heavy elments will be needed for an older planet with a radius at the top of error bar with a core composed of rock.  We find that a range of core masses from 59 and 92 \me~fall within these bounds.  If the age is locked at 2 Gyr and the radius at the value at the center of the error bars (0.726 \rj), these estimates change to 66 \me~(for rock) and 79 \me~(for ice).  Including the $\sim$1-2.5 \me~of heavy elements in the envelope ($Z$=0.045), we derive a total mass of heavy elements for the most disparate models of of 61.5 and 93$\,$\me, which are 54 and 82 \% of the planet's mass, respectively.  The contraction of the radius of two planet models are shown in \mbox{Figure~\ref{figure:RvsT}}.  The radius of a solar composition planet (with the same atmosphere) would be 63\% larger.

As a check on these results, we also investigated the contraction of \hh~with an alternative ``rock'' and ``ice'' EOS, which are given in \citet{Hubbard89}.  These EOSs are polynomial fits to high pressure EOSs given in \citet{ZT78}.  ``Rock'' is a mixture of 38\% SiO$_2$, 25\% MgO, 25\% FeS, and 12\% FeO.  This material is slightly denser at high pressure than our ANEOS rock EOS, leading to a total heavy element mass smaller by $\sim$~1 \me~.  ``Ice'' here is a mixture of 56.5\% H$_2$O, 32.5\% CH$_4$, and 11\% NH$_3$ at temperatures of $\sim$10$^4$ K.  Again we find a total heavy element mass smaller by 1 \me.  Our conclusions regarding heavy elements remain essentially unchanged despite using a different composition for the rock and a non-zero temperature EOS for the ice mixture.


The evolution models show that the interior of \hh~has cooled significantly, which is not surprising given the relatively small percentage of its mass contributed by H/He.  We find that values of $T_{\mathrm{int}}$ from 80-105 K are realistic for the current intrinsic heat flux of the planet.  For comparison, the $T_{\mathrm{int}}$ for our solar system's giant planets are 99 K for Jupiter, 77 K for Saturn, 53 K for Neptune, and $<35$ K (1$\sigma$ upper limit) for Uranus \citep{Pearl91}.  The $T_{\mathrm{int}}$ for \hd~may be as high as 300 K, in order to match the planet's current radius \citep{Guillot02,Baraffe03}.  We also note that if observations show that the planet does indeed have a hot stratosphere, it may be necessary to revisit these evolutionary calculations.  If stellar flux is absorbed very high in the planet's atmosphere this will lessen somewhat the effect that stellar irradiation has on retarding the contraction of the planet \cp{Guillot02}.

\subsection{Comparative Interior Structure}
\begin{figure}
\plotone{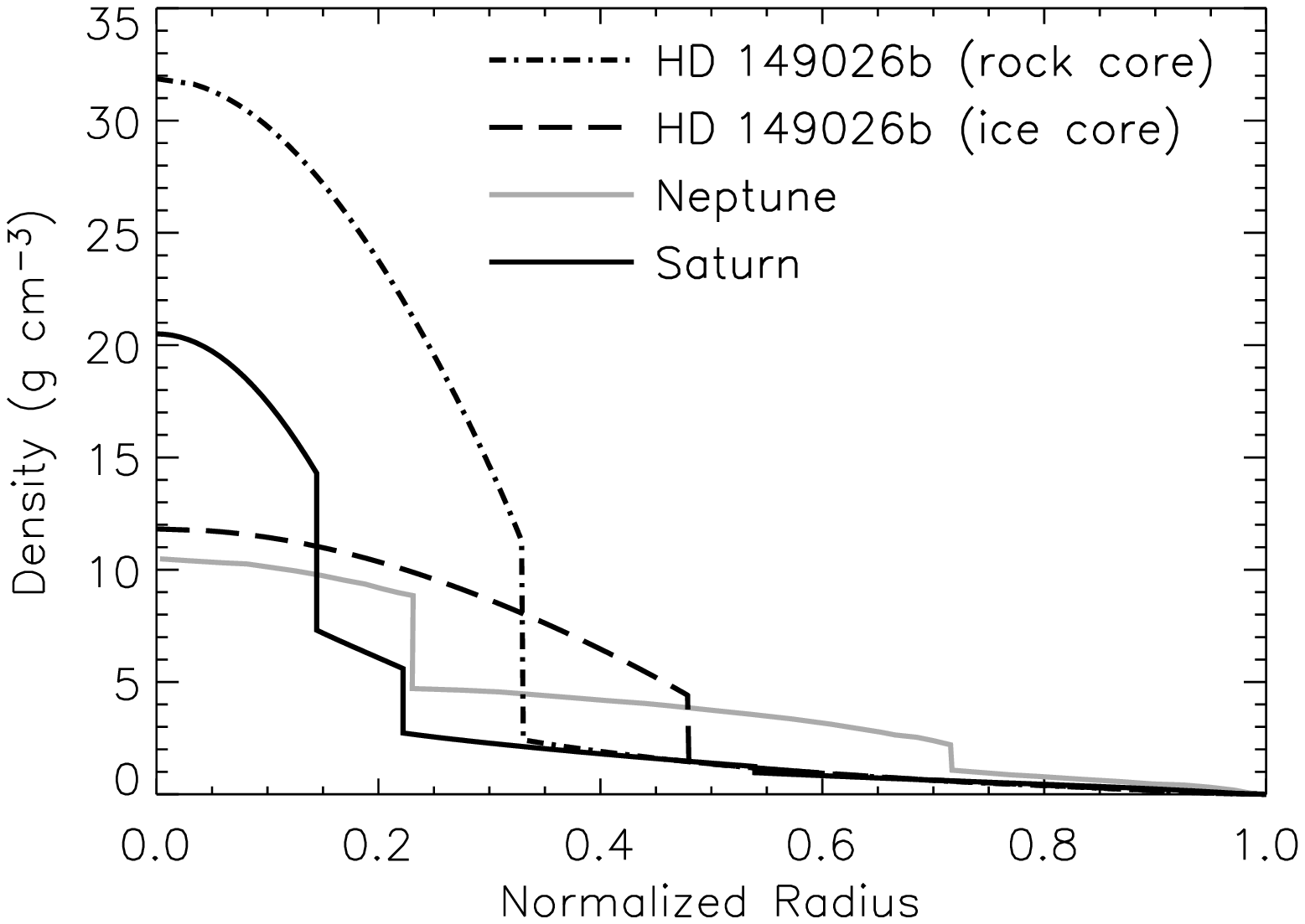}
\caption{Interior density as a function of normalized radius for two possible models for \hh~compared with Neptune and Saturn.  All planet models have been normalized to the radius at which $P$=1 bar.  The Neptune profile is from \citet{Podolak95} and the Saturn profile is from \citet{Guillot99}.  The Saturn and Neptune models have a layer of ice overlying a central rocky core.  The two profiles of \hh~assume that [M/H]=0.5 in the H/He envelope and a core made entirely of either ice or rock.
\label{figure:int}}
\end{figure}

\mbox{Figure~\ref{figure:int}} shows the current interior density distribution as a function of normalized radius for our two models from \mbox{Figure~\ref{figure:RvsT}} compared to interior models of Saturn \citep{Guillot99} and Neptune \citep{Podolak95}.  The Saturn and Neptune models both have two-layer cores of rock overlayed by ice.  The ratio of ice to rock in these cores is based more on cosmogonical arguments than on physical evidence.  The interior structure of \hh~may be a hybrid of the ice giants and gas giants.  Uranus and Neptune are $\sim$90\% heavy elements, while Saturn is $\sim$25\% and Jupiter $\wig<$10\% \citep{Saumon04}.  Although \hh~is more massive than Saturn, it has a bulk percentage of heavy elements (50-80\%) more similar to that of the solar system's ice giants.

For the models in \mbox{Figure~\ref{figure:int}} we find that the central pressure for the planet with the ice core is 72 Mbar, while with the rock core it is 240 Mbar.  At the bottom of the H/He envelope the pressure is 5.6 Mbar in the model with an ice core and 16 Mbar in the model with a rock core. Under these conditions, hydrogen is expected to be compressed into a liquid metallic state.  The possibility of a relatively thin liquid metallic hydrogen layer overlying a deeper region of fluid ionic ices could lead to a interesting new regime of planetary magnetic fields.  Convecting fluid ices overlying non-convective regions may be responsible for the complex magnetic fields of Uranus and Neptune \citep{Ubook,Nbook,Stanley04}.  The geometry of the convective region of a planet's interior may well be an important factor in determining the morphology of a planet's magnetic field \citep{Stanley04}.

\subsection{Composition and the Radii of Pegasi Planets}
The radii of Pegasi planets are influenced by a number of factors.  These include the age of the planet, the magnitude of the stellar flux incident on the planet, the opacity of the atmosphere (which regulates interior cooling), the time delay between formation and the end of migration, and the interior heavy elements abundances in the H/He envelope and core.  The discovery of \hh~has shown us that interior heavy element abundances of Pegasi planets may be quite different than that of Jupiter and Saturn.  \citet{Guillot05} points out that for Pegasi planets in general, the orbital velocity ($v_{\rm orb}$) of $\sim$150 km s$^{-1}$ is much larger than the planet's escape velocity ($v_{\rm esc}$) of $\sim$50 km s$^{-1}$.  As discussed in \citet{Guillot00}, this means Pegasi planets will be relatively inefficient at scattering planetesimals near its orbit, and relatively efficient at capturing them, compared to Jupiter.  Jupiter, at its current mass and location, has $v_{\rm orb} < v_{\rm esc}$ and is efficient at removing planetesimals from the solar system.  Thus, Pegasi planets may accumulate more heavy elements than has been previously appreciated \citep{Guillot05}.  \hh~may be towards the high-accumulation end of a continuum for Pegasi planets, possibly due to the high metallicity in the stellar system it occupies.

This leads us to consider the bulk content of heavy elements of the other transiting planets.  \hd~is the only known transiting planet with a large radius that is difficult to model, although OGLE-TR-10b may be in this class \citep{Konacki05,Baraffe05}.  This has suggested to many that this \emph{one} planet harbors an additional energy source that slows it contraction \citep{Bodenheimer01,Guillot02,Baraffe03,OblTides}.  The discovery of \hh~suggests another possibility.  It is conceivable that \emph{all} Pegasi planets are affected by an additional energy source (yet to be definitively identified) as they all share similar environments.  An internal energy source would normally tend to lead to larger planetary radii, but this could be partly compensated by a relatively large bulk heavy element abundance in these planets, thanks to favorable conditions for planetesimal accretion.

In this perspective, \hd~and OGLE-TR-10b would represent the low end of the metallicity range of the Pegasi planets, leading to relatively large radii.  \hh~would fall at the high metallicity end of of this range, leading to its small radius.  Given Jupiter and Saturn's large bulk inventory of heavy elements---as much as 6 times and 14 times solar, respectively \citep{Saumon04}---and the fact that most calculations on the radii of transiting planets have assumed metallicities lower than that of Jupiter \citep[e.g.][]{Chabrier04,Burrows04,Guillot05,Baraffe05}, this scenario is quite plausible.\footnote{\citet{Bodenheimer03} and \citet{Laughlin05} have computed evolution models with 40 \me~cores for planets more massive than 1 \mj.}  A similar argument was made by \citet{Guillot05}.  The detection of additional transiting planets with similar orbital characteristics but with a continued wide spread in radius would support this view.  In tentative support of this picture, \citet{Fortney05} found that the match to the \citet{Charb05} \emph{Spitzer} photometry for the transiting planet \T~improved as the metallicity of the model atmosphere was increased.  It is tempting to look for trends in the metallicity of the transiting planet host stars as as a hint to potential core sizes in these planets, given the supersolar abundances of HD 149026. However, stars HD 209458, OGLE-TR-10, and \T are all consistent with solar abundances \citep{Brown01,Konacki05,Sozzetti04}, although the planet radii differ by $\sim$0.3 \rj~at similar planetary masses.  Planetary formation details surely complicate this picture, but perhaps as more planets are found a rough trend will emerge.

\section{Conclusions}
We have investigated aspects of the atmosphere, interior, and evolution of the new transiting planet \hh.  If the determination of the planet's radius withstands additional scrutiny then this planet is unlike any other found to date.  We confirm that \hh~is a Pegasi planet composed of 50-80\% heavy elements.  We note that if we had included the transit radius effect, equations of state for ices and rocks that included the additional thermal energy component, or an additional internal energy source, the planet would need to be even more enriched in heavy elements to match its current radius.

Although we calculated models with up to ten times the solar metallicity, the possibility exists that the atmosphere could be even more enriched in heavy elements.  For example, the atmospheres of Uranus and Neptune are enriched in carbon (in the form of CH$_4$) by factors of $\sim$30 over the solar value \citep{Gautier89}.  Given our trends from [M/H]=0.0 to 1.0, a greater metallicity enhancement would lead to a warmer atmosphere that would be even more likely to have a hot stratosphere.  If the planet was able to accrete significant amounts of planetesimals into its proto-atmosphere in this metal-rich environment, it could contain more heavy elements in its H/He envelope and less in its core than we have presumed, although the effects on the radius would be similar.  A signature of this history would be an extremely metal-rich atmosphere.

Detection of \hh~with \emph{Spitzer} may prove challenging, given that the planet-to-star flux density ratios we predict here are smaller than for \T.  However, the 8.0 $\mu$m flux observed from \T~is larger than our models predict, and if this is also the case for \hd~then the case for observing \hh~will be strengthened.  The atmosphere of \hh~is likely to be highly enriched in heavy elements, in which case it could have a stratosphere that may be detected by its strong emission from 4 to 10 $\mu$m.  If a hot stratosphere exists, the planet may be nearly as bright at \T~in the IRAC 3.6 $\mu$m band.

While the stars HD 209458 and HD 189733 have a metallicity close to solar \citep{Mazeh00,Bouchy05}, their transiting planets may nevertheless have hot stratospheres, perhaps driven by absorption by photochemically derived species.  Additional \emph{Spitzer} IRAC data should help show us the shape of the infrared spectrum from 3 to 10 $\mu$m.  Because their parent stars are bright and these systems have favorable planet-to-star flux ratios, they are promising targets to search for an extrasolar stratosphere.  As every object in our solar system with an appreciable atmosphere exhibits a hot stratosphere, save possibly Venus, finding a stratosphere on some or all Pegasi planets would open an exciting new area of comparative planetary atmospheres.\\

We thank Greg Laughlin for sharing data in advance of publication and 
David Charbonneau for discussions about the IRAC detectors.  A review by Tristan Guillot has improved the paper.  We acknowledge support from an NRC postdoctral fellowship (J.~J.~F.), NASA grants NAG2-6007 and NAG5-8919 (M.~S.~M.), and NSF grant AST-0406963 (K.~L.).  This work was supported in part by the US Department of Energy under contract W-7405-ENG-36.
                                                                                       

\newpage

\newpage
\begin{deluxetable}{cccccccc}
\center
\tablecolumns{8}
\tablewidth{0pc}
\tablecaption{Model Planet-to-Star Flux Density Ratios in \emph{Spitzer} Bands}
\tablehead{
\colhead{Planet} & \colhead{Model} & \colhead{$T_{\mathrm{eff}}$ (K)} & \colhead{3.6 $\mu$m} &
\colhead{4.5 $\mu$m} & \colhead{5.8 $\mu$m} & \colhead{8.0 $\mu$m} & \colhead{24 $\mu$m}}
\startdata 
HD 149026b & 3$\times$,4$\pi$ & 1736 & 0.24 & 0.30 & 0.37 & 0.45 & 0.68\\ 
HD 149026b & 3$\times$,4$\pi$, no clouds & 1734 & 0.24 & 0.28 & 0.35 & 0.44 & 0.66\\ 
HD 149026b & 3$\times$,2$\pi$ & 2148 & 0.43 & 0.60 & 0.71 & 0.85 & 1.11\\ 
HD 149026b & 10$\times$,4$\pi$ & 2025 & 0.37 & 0.55 & 0.66 & 0.82 & 1.09\\
HD 149026b & 10$\times$(1$\times$TiO),4$\pi$ & 1852 & 0.32 & 0.44 & 0.56 & 0.67 & 0.94\\
\\ 
HD 189733b & 1$\times$,4$\pi$  & 1190 & 1.04 & 1.77 & 2.31 & 3.12 & 5.69\\ 
HD 189733b & 3$\times$,4$\pi$  & 1199 & 1.24 & 1.74 & 2.40 & 3.30 & 5.87\\ 
HD 189733b & 10$\times$,4$\pi$ & 1212 & 1.35 & 1.82 & 2.58 & 3.51 & 6.19\\ 
\enddata
\tablecomments{All ratios have been multiplied by 1000.  Abundances and number of steradians for reradiation are given.  }
\end{deluxetable}

\end{document}